\documentclass[sn-mathphys]{sn-jnl}

\jyear{2021}%

\theoremstyle{thmstyleone}%
\theoremstyle{thmstyletwo}%

\theoremstyle{thmstylethree}%

\usepackage{stmaryrd}
\usepackage{proof}
\usepackage{graphicx}
\usepackage{dcolumn}
\usepackage{bm}
\usepackage{qcircuit}
\usepackage{amsmath}  
\usepackage{listings}
\usepackage{xy}
\usepackage{physics}
\usepackage{amsthm}

\begin{document}

\title[Quantum computations for disambiguation and question answering]{Quantum computations for disambiguation and question answering}


\author*[1,3]{\fnm{A. D.} \sur{Correia }}\email{a.duartecorreia@uu.nl}

\author[2,3]{\fnm{M.} \sur{Moortgat}}

\author[1,3]{\fnm{H. T. C.} \sur{Stoof}}

\affil[1]{\orgdiv{Institute for Theoretical Physics}, \orgname{Utrecht University}, \orgaddress{Princetonplein 5}, \city{Utrecht}, \postcode{3584 CC}, \country{The Netherlands}}

\affil[2]{\orgdiv{Utrecht Institute of Linguistics OTS}, \orgname{Utrecht University}, \orgaddress{Trans 10}, \postcode{3512 JK}, \city{Utrecht}, \country{The Netherlands}}

\affil[3]{\orgdiv{Center for Complex Systems Studies}, \orgname{Utrecht University}, \orgaddress{Leuvenlaan 4}, \postcode{3584 CE}, city{Utrecht}, \country{The Netherlands}}


\abstract{Automatic text processing is now a mature discipline in computer science, and so attempts at advancements using quantum computation have emerged as the new frontier, often under the term of Quantum Natural Language Processing. The main challenges consist in finding the most adequate ways of encoding words and their interactions on a quantum computer, considering hardware constraints, as well as building algorithms that take advantage of quantum architectures, so as to show improvement on the performance of natural language tasks.
In this paper, we introduce a new framework that starts from a grammar that can be interpreted by means of tensor contraction, to build word representations as quantum states that serve as input to a quantum algorithm.  We start by introducing an operator measurement to contract the representations of words, resulting in the representation of larger fragments of text. We then go on to develop pipelines for the tasks of sentence-meaning disambiguation and question answering that take advantage of quantum features. For the first task, we show that our contraction scheme deals with syntactically ambiguous phrases storing the various different meanings in quantum superposition, a solution not available on a classical setting. For the second task, we obtain a question representation that contains all possible answers in equal quantum superposition, and we implement Grover's quantum search algorithm to find the correct answer, agnostic to the specific question, an implementation with the potential of delivering a result with quadratic speedup.}




\maketitle

\section{\label{sec:level1}Introduction}

Recent developments in quantum computation have given rise to new and exciting applications in the field of Natural Language Processing (NLP).  Pioneering work in this direction is the DisCoCat framework \cite{coecke2010mathematical,Coecke_2013}, which introduces a compositional mapping between types and derivations of Lambek's typelogical grammars \cite{lambek1958mathematics,Lambek97} and a distributional semantics \cite{turney2010frequency} based on vector spaces, linear maps and tensor products. In this framework, the interpretations of large text fragments are obtained by performing a tensor contraction between the tensor interpretations of individual words. To interpret text fragments taking into account their grammatical features, while staying in the vector space semantics, the dimension of the representation quickly scales, as it depends on the complexity of the syntactic type, which has been a limiting feature in vector-based semantics implementations \cite{wijnholds2020representation}. This motivates a representation of words as quantum states, counting on the potential of quantum computers to outperform the limitations of classical computation both in terms of memory use \cite{giovannetti2008quantum} and in processing efficiency \cite{arute2019quantum}. In this setting, words are represented as multi-partite quantum states, with the theory predicting that, when contracted with one another, the meaning of larger text fragments is encoded in the resulting quantum states. 

The challenge is now in implementing these contractions on quantum circuits. Circumventing this issue, DisCoCirc \cite{coecke2019mathematics} introduces a different way of representing the meaning of a sentence, where certain words are seen as quantum gates that act as operators on input states representing other words. The DisCoCirc approach uses quantum machine learning algorithms \cite{biamonte2017quantum} for NLP \cite{meichanetzidis2020quantum,coecke2020foundations} where circuit parameters, related to word representations, are then learned by classical optimization and used to predict different binary labels statistically, such as the answers to \textit{yes-no} questions \cite{meichanetzidis2020grammar}, topics of phrases, or the distinction between subject and object relative clauses \cite{lorenz2021qnlp}. 

Although these implementations can play an important role in speeding up NLP tasks based on current machine-learning ideas and techniques, they do not go beyond the current paradigm in terms of classification tasks. Furthermore, a number of theoretical advances using the tensor contractions from DisCoCat cannot be directly reproduced, since the mapping from a phrase to a circuit requires extra steps that deviate from the original grammatical foundation, not treating every word as an input at the same level. We refer here to the work done in expanding the toolbox of word representations with density matrices \cite{piedeleu2015open}, so as to achieve good results on discerning different word and phrase senses \cite{DBLP:journals/amai/SadrzadehKB18,bankova2019graded,meyer-lewis-2020-modelling}, and in entertaining simultaneously different possible interpretations of texts, either by looking at an incremental interpretation of the parsing process \cite{shiebler2020incremental}, or by considering a single representation for the multiple readings of syntactic ambiguities \cite{correia2020density,correia2020putting}. This presents a strong incentive to find an alternative quantum-circuit implementation that sticks to the original grammatical formulation, preserving the previous achievements, where all words are taken as input on an equal footing.
In addition, it is our belief that a quantum framework can contribute a great deal to the reestablishment of rule-based NLP, as a desirable alternative to large-scale statistical approaches \cite{bender2021dangers}, since certain computations become more efficient if we use the appropriate quantum algorithms, as we will illustrate in the case of question-answering where quadratic quantum speedup can be achieved.

The paper is structured as follows. In Sec. \ref{syntax} we develop the grammatical framework and quantum state interpretation thereof, setting the stage for the types of linguistic problems we will deal with here. Here we introduce the idea that words are represented as vectors, matrices, and higher-rank tensors, depending on their grammatical function, that contract with each other following grammatical rules, explaining how we can arrive at the interpretations of larger fragments of text.  In Sec. \ref{implementation} we put forward an approach where the words are interpreted as quantum states, and we show how the contractions between word representations can be implemented on a quantum computer as the measurement of a permutation operator. We elaborate on how this setting permits the simultaneous treatment of ambiguous phrases in English. In Sec. \ref{application} we apply Grover's algorithm to question-answering, using the framework developed in the previous section to turn the representation of the question and answers into the input of the algorithm, together with an oracle that identifies that correct answers. Finally, in Sec. \ref{conclusion} we give an overview of the developments introduced and discuss further work.

\section{Syntax-semantics interface}\label{syntax}
In this section we introduce the grammatical framework that we will be working with. It consists of a categorial grammar as the syntactic front end, together with a compositional mapping that sends the types and derivations of the syntax to a vector-based distributional interpretation. This is necessary to understand the type of linguistic problems that we can address and how they can be solved using a quantum circuit. 

\subsection{Type logic as syntax}
The key idea of categorial grammar formalisms is to replace the parts of speech of traditional grammars (nouns, adjectives, (in)transitive verbs, etc) by logical formulas or types; a deductive system for these type formulas then determines their valid combinations. The idea can be traced back to Ajdukiewicz \cite{ajdukiewicz1935syntaktische}, but Lambek's Syntactic Calculus \cite{lambek1958mathematics} is the first full-fledged formulation of a categorial type logic that provides an algorithm to effectively decide whether a phrase is syntactically well formed or not.

Let us briefly discuss types and their combinatorics. We start from a small set of primitive types, for example $s$ for declarative sentences, $n$ for noun phrases, $w$ for open-ended interrogative sentences, etc. From these primitive types, compound types are then built with the aid of three operations: multiplication $\bullet$, left division $\backslash$ and right division $/$. Intuitively, a type $A\bullet B$ stands for the concatenation of a phrase of type $A$ and a phrase of type $B$ (``$A$ and then $B$''). Concatenation is not commutative (``$A$ and then $B$'' $\neq$ ``$B$ and then $A$''). Hence we have left \textit{vs} right division matching the multiplication: $A\backslash B$ can be read as ``give me a phrase $A$ to the left, and I'll return a phrase $B$''; $B/A$ is to be interpreted as ``give me a phrase $A$ to the right, and I'll return a phrase $B$''. We can codify this informal interpretation in the rules below, where $A_1 \bullet \cdots \bullet A_n \vdash B$ means that from the concatenation of phrases of type $A_1,\ldots,A_n$ one can derive a phrase of type $B$. Hence,

\begin{align}
&B\slash A \bullet  A  \vdash  B  \label{contright}, \\
&  A \bullet A \backslash B   \vdash B. \label{contleft}
\end{align}

As examples of simple declarative sentences, consider \emph{Alice talks}, or \emph{Bob listens}. In the former case, we assign the type $n$ to \emph{Alice} and the type $n\backslash s$ to the intransitive verb \emph{talks}. We start by multiplying the word types in the order by which the words appear, forming $n \bullet n\backslash s$. Then, it suffices to apply rule (\ref{contleft}), with $A=n$ and $B=s$, to show that $n \bullet n\backslash s$ derives $s$, i.e.~constitutes a well-formed sentence. Conversely, the lack of a derivation of $s$ from $n\backslash s \bullet n$ (\emph{talks Alice}) allows us to conclude that this not a well-formed sentence. These, and the later examples, illustrate only the simplest ways of combining types, but these will suffice for the purposes of this paper. To obtain a deductive system that is sound and complete with respect to the intended interpretation of the type-forming operations, Lambek's Syntactic Calculus also includes rules that allow one to infer $A\vdash C/B$ and $B\vdash A\backslash C$ from $A\bullet B\vdash C$. Moreover, to deal with linguistic phenomena that go beyond simple concatenation, Lambek's type logic has been extended in a number of ways that keep the basic mathematical structure intact but provide extra type-forming operations for a finer control over the process of grammatical composition. See Ref. \cite{moortgat1997categorial} for a survey, and Ref. \cite{correia2020putting} for a quantum interpretation of such structural control operations.

\subsubsection{Syntactic ambiguities}

To see rule (\ref{contright}) in action, consider adjectives in English. An adjective is expecting a noun to its right, and, once it is composed with a noun, it must derive something that can be used, for instance, as the argument of an intransitive verb, which, as we have seen, is of type $n$. Thus, an adjective must be of type $n\slash n$, and we can use rule (\ref{contright}) to prove that, as an example, \textit{rigorous mathematicians} is a well-formed phrase of type $n$. 

For certain phrases, there is more than one way of deriving the target type, with each derivation corresponding to a distinct interpretation. As an example, consider the noun phrase \textit{rigorous mathematicians and physicists}, an ambiguous structure that has already been studied in the context of vector representations in Ref. \cite{correia2020density}. Here the conjunction \textit{and} gets the type $(n\backslash n) \slash n$; for the complete phrase, we want to show that the following judgement holds:

\begin{equation}
    n \slash n \bullet n \bullet (n\backslash n) \slash n \bullet n \vdash n.
\end{equation} There are two possible interpretations: a first one, where the adjective \textit{rigorous} has scope over \textit{mathematicians and physicists}, and a second one, where it only has scope over \textit{mathematicians}. Each of these interpretations is connected to a different way of deriving the goal formula $n$. The first reading is obtained by applying the rules in the following order

\begin{equation}\label{firstreading}
    \underbrace{n \slash n \bullet \underbrace{n \bullet \underbrace{(n\backslash n) \slash n \bullet n}_{\text{(\ref{contright}) } \vdash n\backslash n}}_{\text{(\ref{contleft}) } \vdash n }}_{\text{(\ref{contright}) } \vdash n },
\end{equation} while for the second reading the rules apply in a different order as

\begin{equation}\label{secondreading}
    \underbrace{\underbrace{n \slash n \bullet n}_{\text{(\ref{contright}) } \vdash n} \bullet \underbrace{(n\backslash n) \slash n \bullet n}_{\text{(\ref{contright}) }\vdash n\backslash n}}_{\text{(\ref{contleft}) } \vdash n }.
\end{equation} Our goal is to treat both readings simultaneously until further information allows us to clarify which of the readings is the intended one.

\subsubsection{Question answering}

Question answering (Q$\&$A) is one of the most common tasks in NLP \cite{soares2020literature}. Questions can be close ended, having ``yes" or ``no" for an answer, or open ended, starting by ``who", ``why" or ``what", also referred to as \textit{wh}-questions. For $P$ possible answers, it is always possible to turn \textit{wh}-questions into close-ended questions. If we know that either Alice, Bob, Carol or Dave is talking, we can turn ``Who talks?" into a series of four questions ``Does  \texttt{[name]}  talk?". Thus, for $P$ possible answers, there are $P$ closed-ended questions that we need to check \footnote{This is a common way of turning Q$\&$A into a classification problem, where each close-ended question gets a binary label, depending on whether the answer is true or false. Binary classification problems are
some of the most well established applications of machine learning. After finding a way of representing the question statements, usually as single vectors, a number of these labeled statements is used to predict the labels of the hold-out statements.}. We would like to find the answer to the open-ended questions directly, without this mapping. Syntactically,
\textit{wh}-questions are open-endend interrogative sentences, and as such are assigned their own type $w$. For a subject question, the type of the word \textit{who} is thus $w \slash (n\backslash s)$, since, when applied to an intransitive verb using rule (\ref{contleft}), it derives the interrogative type $w$.

\subsection{Vectors as semantics}

In the context of automatic processing of text, the most widely used form of representing a word is by a unique array of values, referred to as a ``word embedding". Seen as vectors, we can cluster or compare them using varied geometric tools \cite{lin1998automatic,navigli2010inducing,nasiruddin2013state}. Representing the meanings of words as such is widely known as ``distributional semantics" \cite{boleda2020distributional}. In earlier work, vector entries were related with how often a word would appear next to other words \cite{rieger1991distributed}, following the ``distributional hypothesis" that states that words that appear in similar contexts are themselves similar \cite{harris1954distributional}. Nowadays, word embeddings are extracted using language models, targeted on the prediction of the most likely next word \cite{mikolov2013distributed,devlin2018bert}. This presents a problem for the representation of larger fragments, since they are less likely to appear in a text, making their distributional array rather sparse and thus not particularly meaningful. Larger fragments can nevertheless receive an embedding, but a direct connection with grammatical composition is lost.

To tackle this problem, the authors in Ref. \cite{coecke2010mathematical} propose that the arrays representing different words depend on their syntactic types, namely having a dimensionality that mirrors their type complexity. This introduces a way of composing the meanings of the individual words that is homomorphic to the syntactic derivations, generating a representation of larger fragments from the representation of smaller ones. For completeness, the mapping between the syntax and the semantics is done using the formalism of vector spaces. Each syntactic type $A$ is mapped to its semantic type via $\lceil A \rceil$. Each semantic type is then interpreted as a vector space, where the particular words are represented.
Let there be three basic semantic spaces $\{S,N,I\}$. The simple syntactic types $n$ and $s$ are mapped respectively to $\lceil n \rceil = N$ and $\lceil s \rceil = S$. Each individual word is an element of the semantic space that interprets its syntactic type. For instance, the interpretation of the word \textit{physicists} is now seen as a vector in $N$, this being the vector space where the distributional information of nouns is stored. Similarly, \textit{Alice talks} is represented by a vector in $S$, that has as basis elements two orthogonal states corresponding to ``true" and ``false". The interrogative type $w$ is mapped to $\lceil w \rceil = I \otimes N \otimes I \otimes S$. The vector space $I$ (``index") has basis elements that are in one-to-one correspondence to the nouns that can be used as answers to the interrogative sentence, providing an enumeration of the noun vectors of $N$. This will be useful later when we need to index the quantum states associated with each possible answer.

The vector spaces that translate the directional and multiplicative types are obtained recursively as 
\begin{align}
&    \lceil A \backslash B \rceil = \lceil A \slash B \rceil = \lceil A \bullet B \rceil = \lceil A \rceil \otimes \lceil B \rceil,
\end{align} where $\otimes$ forms a tensor product space, inductively starting from $A,B,C \in \{n,s,w\}$. Note that the tensor is commutative, such that $\lceil A \rceil \otimes \lceil B \rceil \cong \lceil B \rceil \otimes \lceil A \rceil$.  We perform tensor contractions as the interpretations of the rules in Eqs. (\ref{contright}) and (\ref{contleft}). Thus, an intransitive verb is represented as a matrix in $N \otimes S$, that when acting on a vector of type $N$ returns a vector in $S$. Using the notation $\llbracket . \rrbracket$ to represent the tensor interpretation of a word, and assuming an orthogonal basis $\{\hat{n}_i\}$ of $N$ and an orthogonal basis $\{\hat{s}_i\}$ of $S$, the composition of the vectorial interpretations of \textit{Alice} and \textit{talks} leads to the interpretation of the entire sentence as a vector in $S$. The word meanings for this sentence are represented as

\begin{align}
&\llbracket \text{Alice} \rrbracket= \sum_p A_p \; \hat{n}_p \\
 & \llbracket \text{talks} \rrbracket= \sum_{qr} t_{qr} \; \hat{n}_q \otimes \hat{s}_r,
\end{align} and the full sentence meaning as

\begin{align}\label{aliceflies1}
&\llbracket \text{Alice talks} \rrbracket = \llbracket \text{Alice} \rrbracket \cdot \llbracket \text{talks} \rrbracket = \sum_{pr} A_p t_{pr} \hat{s}_r .
\end{align} A more refined treatment of the translation from the Lambek types to tensor spaces has been given in Ref. \cite{correia2020density}.

Similarly, the semantic space for an adjective can be seen as a matrix in $N \otimes N$. Note that here the analogy between a matrix modifying a vector and the adjective as a noun modifier is the clearest. 
Let us look at the meanings of \textit{rigorous} and \textit{mathematicians}, which can be represented as

\begin{align}
 & \llbracket \text{rigorous} \rrbracket= \sum_{ij} r_{ij} \; \hat{n}_i \otimes \hat{n}_j \\
&\llbracket \text{mathematicians} \rrbracket= \sum_k m_k \; \hat{n}_k.
\end{align} The meaning of \textit{rigorous mathematicians} will be given by the application of the translation of rule (\ref{contright}) to tensors. At the components level, it is the matrix multiplication between the \textit{rigorous} matrix and the \textit{mathematicians} vector, which gives, consistently with $n$ being the syntactic type of this fragment, a vector in $N$, as

\begin{align}
&\llbracket \text{rigorous} \; \text{mathematicians} \rrbracket \nonumber \\
&= \llbracket \text{rigorous} \rrbracket . \llbracket \text{mathematicians} \rrbracket =\sum_{ij} r_{ij} \, m_j \; \hat{n}_i. 
\end{align} The different order of application of Lambek rules in Eqs. (\ref{firstreading}) and (\ref{secondreading}) translates into different vectors that represent the two readings of \textit{rigorous mathematicians and physicists}. The words \textit{and} and \textit{physicists} are given the vector representations

\begin{align}
&\llbracket \text{and} \rrbracket= \sum_{lmn} a_{lmn} \; \hat{n}_l \otimes \hat{n}_m \otimes \hat{n}_n, \\
& \llbracket \text{physicists} \rrbracket = \sum_{o} p_{o} \; \hat{n}_o. 
\end{align} The reading from Eq. (\ref{firstreading}) is represented by the vector

\begin{align}\label{firstreadingfinal}
 &   \llbracket \text{rigorous mathematicians and physicists} \rrbracket_1 = \sum_{ijln} \, r_{ij} \, m_{l} \, a_{ljn} \, p_{n} \; \hat{n}_i,
\end{align} whereas the reading from Eq. (\ref{secondreading}) is encoded in the vector 
\begin{align}\label{secondreadingfinal}
&   \llbracket \text{rigorous mathematicians and physicists} \rrbracket_2    = \sum_{jlmn} \, r_{lj} \, m_{j} \, a_{lmn} \, p_{n} \; \hat{n}_m,
\end{align} which are of the same form as the results in Ref. \cite{correia2020density}.

For interrogative sentences, the word \textit{who} will have the semantic function of ``lifting" an intransitive verb with representation in space $N\otimes S$ to a representation in $I \otimes N \otimes I \otimes S$, since
\begin{align}\label{whoref} 
 &   \lceil w \slash (n\backslash s) \rceil = I \otimes N \otimes I \otimes S \otimes N\otimes S \nonumber \\
 & \cong I \otimes N \otimes I \otimes S \otimes S \otimes N.
\end{align} An element of this space contracts with an element of the representation space of intransitive verbs, $N\otimes S$, associating the index of every possible answer, in $I$, with both its representation in $N$ and its truth value in $S$.

\section{Implementation}\label{implementation}

In this section we motivate a passage from vectors to quantum states and we introduce them as inputs of quantum circuits that calculate contractions between word representations.

\subsection{Quantum states as inputs of a quantum circuit}

We now switch to a representation of word embeddings as vectors in complex-valued inner product vector spaces, i.e. Hilbert spaces. Our atomic semantic spaces $N$, $S$ and $I$ will now be replaced by their quantum counterparts as the interpretation spaces. We thus have the Hilbert spaces $\mathcal{H}^N$, $\mathcal{H}^S$ and $\mathcal{H}^{\otimes p}$ respectively, with $\mathcal{H}^{\otimes p}$ the $p$-qubit Hilbert space corresponding to the complex-valued realization of the semantic type $I$, where we assume that $P=2^p$. For instance, with $\{ \ket{n_i} \}$ the basis of $\mathcal{H}^N$, we now have

\begin{equation}
    \llbracket \text{Alice} \rrbracket = \ket{\text{Alice}} = \sum_{p} A_{p} \; \ket{n_p}. 
\end{equation} Note that this space allows us to expand our representations with complex-valued entries, and a proper contraction between the words will require the conjugation of some of the components, i.e.

\begin{equation}
    \llbracket \text{Alice} \rrbracket^* = \bra{\text{Alice}} = \sum_{p} A^*_{p} \; \bra{n_p}. 
\end{equation} 

Let the input of a circuit be the product of the states that interpret each word in the language fragment in question. Our running example of a noun subject and an intransitive verb \textit{Alice talks} is now represented as the input
\begin{equation*}
\ket{\text{Alice}} \ket{\text{talks}} \in \mathcal{H}^N \otimes \mathcal{H}^N \otimes \mathcal{H}^S. 
\end{equation*} The basis of $\mathcal{H}^S$, $\{\ket{s_i}\}$, are the single-qubit spin states $\ket{0}$ and $\ket{1}$, where the former represents a sentence that is false, and the later one that is true. In this setting, it is also possible to establish a probability distribution over the truthfullness of a sentence.

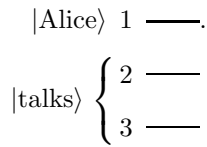
\begin{figure}[!tbh]
\[\Qcircuit @C=1em @R=2em @!R {
     & \lstick{\ket{\text{Alice}} \; 1} & \qw & \qw \\
     & \lstick{2} & \qw & \qw \\
     & \lstick{3} & \qw & \qw \inputgroupv{2}{3}{1.2em}{1em}{\ket{\text{talks}} \qquad}
}.\]
\caption{Quantum circuit with intransitive sentence input.}\label{circ1}
\end{figure}

Each of the elements of the interpreting spaces will be represented by a labeled quantum wire, thus rewriting the input state as

\begin{equation}\label{alicefliesinput}
\ket{\text{Alice}} \ket{\text{talks}} \in \mathcal{H}_N^1 \otimes \mathcal{H}_N^2 \otimes \mathcal{H}_S^3, 
\end{equation} used as the input of a quantum circuit, as shown in Fig. \ref{circ1}.

\begin{figure}[!tbh]
\[\Qcircuit @C=1em @R=2em @!R {
     & \lstick{1} & \qw & \qw \\
     & \lstick{2} & \qw & \qw \inputgroupv{1}{2}{1.2em}{1em}{\ket{\text{rigorous}} \; \qquad \quad} \\
     & \lstick{\ket{\text{mathematicians}} \; 3} & \qw & \qw \\
     & \lstick{4} & \qw & \qw \\
     & \lstick{5} & \qw & \qw \\
     & \lstick{6} & \qw & \qw \inputgroupv{4}{6}{1.2em}{2em}{\ket{\text{and}} \; \quad } \\
     & \lstick{\ket{\text{physicists}} \;  7} & \qw & \qw 
}\]
\caption{Quantum circuit with syntactically ambiguous input.}
\label{circ2}
\end{figure}
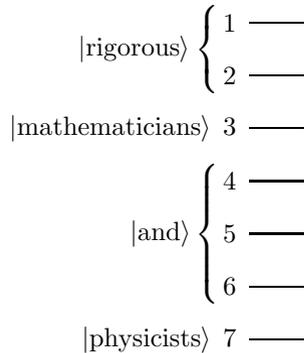

\noindent The ambiguous fragment \textit{rigorous mathematicians and physicists} will be initially represented as a unique state in the tensor space, formed by numbered copies of the $\mathcal{H}^N$ space as

\begin{align*}
&\ket{\text{rigorous}} \ket{\text{physicists}} \ket{\text{and}} \ket{\text{mathematicians}} \\
&\in \mathcal{H}^N_1 \otimes \mathcal{H}^N_2 \otimes \mathcal{H}^N_3 \otimes \mathcal{H}^N_4 \otimes \mathcal{H}^N_5 \otimes \mathcal{H}^N_6 \otimes \mathcal{H}^N_7, 
\end{align*} forming the input of a quantum circuit as in Fig. \ref{circ2}.

\noindent From here, different contractions, corresponding to the two possible readings, will be represented by different circuits acting on this same input, as we show in detail below.

\subsection{Contraction as measurement of permutation operator}\label{demonstration}

To compute the desired contraction using the quantum circuit, we calculate the expectation value of the permutation operator $\widehat{P}_{ij}$ on the input states/wires indexed by $i,j$. These correspond to the spaces with elements that we want to contract, following the syntactic rules. For two states $\ket{\phi_1}_i$ and $\ket{\phi_2}_j$ belonging to two numbered copies of a Hilbert space, respectively $\mathcal{H}^{\lceil A \rceil}_i$ and $\mathcal{H}^{\lceil A \rceil}_j$, we refer to the following quantity as the \textit{measurement of the expectation value of the permutation operator}:

\begin{align}
&    \bra{\phi_1}_i\bra{\phi_2}_j \widehat{P}_{ij} \ket{\phi_1}_i\ket{\phi_2}_j = \nonumber \\
&  = \braket{\phi_1}{\phi_2}_i \braket{\phi_2}{\phi_1}_j \equiv \abs{ \braket{\phi_1}{\phi_2}} ^2.
\end{align} In general, to obtain this quantity on a quantum circuit, one must perform repeated measurements of the input states $\ket{\phi_1}_i$ and $\ket{\phi_2}_j$, on a basis that diagonalizes the permutation operator, summing the frequency of outcomes, using the respective operator's eigenvalues as weights. We introduce the following circuit notation to indicate the measurement of the permutation operator:

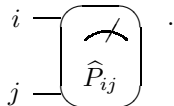
\[\Qcircuit @C=1em @R=2em @!R {
     & \lstick{i} & \multimeasure{1}{\parbox{0.7cm}{\centering $\xy ="j","j"-<.778em,.322em>;{"j"+<.778em,-.322em> \ellipse ur,_{}},"j"-<0em,.4em>;p+<.5em,.9em> **\dir{-},"j"+<1.4em,0.2em>*{},"j"-<1em,1em>*{} \endxy$\\$\widehat{P}_{ij}$}} &  \\
     & \lstick{j} & \ghost{\parbox{0.7cm}{\centering $\xy ="j","j"-<.778em,.322em>;{"j"+<.778em,-.322em> \ellipse ur,_{}},"j"-<0em,.4em>;p+<.5em,.9em> **\dir{-},"j"+<1.4em,0.2em>*{},"j"-<1em,1em>*{} \endxy$\\$\widehat{P}_{ij}$}} &
}. \] If the measuring device is only capable of performing measurements in the standard basis, then we must additionally apply to the input states the inverse of the transformation that diagonalizes the permutation operator, before performing the repeated measurements. In appendix \ref{permutationappendix} we show how this can be achieved using the inverse transformation to the Bell basis in the case of two-qubit inputs. In this case, the measurement of the expectation value can be understood as the map between the SWAP operator, that represents the permutation operator in that case,
and the projection operator on the maximally entangled state $\ket{\beta_{00}}$, which, although not a homomorphism, will be diagonal in the same basis, since both operators share an algebra.

We now show in two ways that the final representation of a simple sentence such as \textit{Alice talks} is stored as an effective state $\ket{\psi}$ in $\mathcal{H}^S_3$, after measuring $\widehat{P}_{12}$ not normalizing for clarity, with input as given in Eq. (\ref{alicefliesinput}).

\paragraph{Using operators.} Assume that an operator $\widehat{O}_3$ is being measured in space $\mathcal{H}^S_3$. Its expectation value is given by $\bra{\psi} \widehat{O}_3 \ket{\psi}$, after measuring $\widehat{P}_{12}$, with

\begin{align}
& \bra{\text{Alice}} \bra{\text{talks}} \widehat{P}_{12} \otimes \widehat{O}_3 \ket{\text{Alice}} \ket{\text{talks}} \nonumber \\
& \equiv \bra{\psi} \widehat{O}_3 \ket{\psi}.
\end{align} The left-hand side unfolds as follows:

\begin{align}\label{alicefliescalc}
&  \bra{\text{Alice}} \bra{\text{talks}} \; \widehat{P}_{12} \otimes \widehat{O}_3 \; \ket{\text{Alice}} \ket{\text{talks}}  = \nonumber \\  
& = \sum_{pqr, p'q'r'} A^*_p t^*_{qr} \bra{n_p n_q s_r} \; \widehat{P}_{12} \otimes \widehat{O}_3 \nonumber \\
& = \;  A_{p'} t_{q'r'} \ket{n_{p'} n_{q'} s_{r'}} \nonumber \\
& =  \sum_{pqr, p'q'r'} A^*_p t^*_{qr} \bra{n_p n_q s_r} \; \widehat{O}_3 \; A_{p'} t_{q'r'} \ket{n_{q'} n_{p'} s_{r'}} \nonumber \\
& =  \sum_{pqr, p'q'r'} A^*_p t^*_{qr} \bra{s_r} \; \widehat{O}_3 \; A_{p'} t_{q'r'} \ket{s_{r'}} \delta_{pq'} \delta_{qp'} \nonumber \\
& =  \sum_{qr, q'r'} A_{q} t^*_{qr} \bra{s_r} \; \widehat{O}_3 \;  A^*_{q'} t_{q'r'} \ket{s_{r'}}.
\end{align}  To uniquely determine $\ket{\psi}$, we need to solve $m=d(d-1)\slash 2$ independent equations, where $d$ is the dimension of $\mathcal{H}^3$, of the form below

\begin{equation}\label{psival}
\sum_{qr, q'r'} A_{q} t^*_{qr} \bra{s_r} \; \widehat{O}_3 \;  A^*_{q'} t_{q'r'} \ket{s_{r'}} =  \bra{\psi} \widehat{O}_3 \ket{\psi}.
\end{equation} Any operator $\widehat{O}_3$ can be decomposed as a sum of $m$ linearly independent operators $\widehat{O}^a_3$, with $1<a<m$. Since Eq. (\ref{psival}) holds for any operator, then it holds for each $\widehat{O}^a_3$, thus generating $m$ independent equations, necessary and sufficient to solve for $\ket{\psi}$. In particular, if $\ket{\psi}$ is expressed in the basis $\ket{s_{r'}}$, the components of $\ket{\psi}$ are given precisely by the respective components of the left-hand side of Eq. (\ref{psival}), from which we can immediately conclude that the effective state in $\mathcal{H}_S^3$ is

\begin{equation}\label{alicefliesfinal}
 \ket{\psi} = \sum_{q'r'} A^*_{q'} t_{q'r'} \ket{s_{r'}} \equiv \llbracket \text{Alice talks} \rrbracket. \qquad \blacksquare
\end{equation} 

Similarly, density matrices can be used to confirm not only that the state in $\mathcal{H}^S_3$ corresponds to Eq. (\ref{alicefliesfinal}) after the measurement, using partial tracing, but also that the outcome of this operation is a pure state.
\paragraph{Using density matrices.} Assume that the sentence \textit{Alice talks} is being represented by the pure state density matrix

\begin{equation}
    \widehat{\rho} = \ket{\text{Alice}}\ket{\text{talks}}\bra{\text{Alice}}\bra{\text{talks}}. 
\end{equation} We want to show that the density matrix $\widehat{\rho}_3$ that we obtain in space $\mathcal{H}_3^{S}$ after the measurement of $\widehat{P}_{12}$ is in fact a pure state. We do that by taking the partial trace in spaces $1$ and $2$ of $\widehat{P}_{12} \widehat{\rho}$:

\begin{align}
& \widehat{\rho}_3 = \Tr_{12} \left( \widehat{P}_{12} \widehat{\rho} \right) = \nonumber \\
& = \sum_{ab} \sum_{pqr, p'q'r'}  \bra{n_a n_b} \widehat{P}_{12} A_{p'} t_{q'r'} \ket{n_{p'} n_{q'} s_{r'}}  \bra{n_p n_q s_r} A^*_p t^*_{qr} \ket{n_a n_b} \nonumber \\
& = \sum_{ab} \sum_{pqr, p'q'r'}  \bra{n_a n_b} A_{p'} t_{q'r'} \ket{n_{q'} n_{p'} s_{r'}} \bra{n_p n_q s_r} A^*_p t^*_{qr} \ket{n_a n_b} \nonumber \\
& =  \sum_{r, p'q'r'}  A_{p'} t_{q'r'} \ket{s_{r'}} \bra{s_r} A^*_{q'} t^*_{p'r} \nonumber \\
& =   \sum_{q'r'} A^*_{q'} t_{q'r'} \ket{s_{r'}} \sum_{p'r} \bra{s_r} A_{p'} t^*_{p'r} = \ket{\psi}\bra{\psi}.
\end{align} This thus proves that the resulting state in space $\mathcal{H}_3^{S}$ is pure and equal to Eq. \ref{alicefliesfinal}. It also proves that $\braket{\psi} = \langle \widehat{P}_{12} \rangle$, as expected from $\bra{\psi} \widehat{O} \ket{\psi} = \langle \widehat{O} \rangle \langle \widehat{P}_{12}  \rangle$. $\blacksquare$

\begin{figure}[!tbh]
\[\Qcircuit @C=1em @R=2em @!R {
     & \lstick{\ket{\text{Alice}} \; 1} & \multimeasure{1}{\parbox{0.7cm}{\centering $\xy ="j","j"-<.778em,.322em>;{"j"+<.778em,-.322em> \ellipse ur,_{}},"j"-<0em,.4em>;p+<.5em,.9em> **\dir{-},"j"+<1.4em,0.2em>*{},"j"-<1em,1em>*{} \endxy$\\$\widehat{P}_{12}$}} &  \\
     & \lstick{2} & \ghost{\parbox{0.7cm}{\centering $\xy ="j","j"-<.778em,.322em>;{"j"+<.778em,-.322em> \ellipse ur,_{}},"j"-<0em,.4em>;p+<.5em,.9em> **\dir{-},"j"+<1.4em,0.2em>*{},"j"-<1em,1em>*{} \endxy$\\$\widehat{P}_{12}$}} & \\
     & \lstick{3} & \qw & \qw \inputgroupv{2}{3}{1.2em}{1.5em}{\ket{\text{talks}} \qquad}.
}\] 
\caption{Quantum circuit that measures the permutation operator $\widehat{P}_{12}$ on an intransitive sentence input.} \label{circ3}
\end{figure}
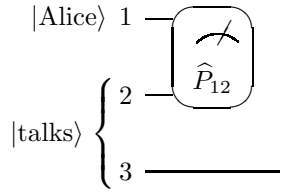

Note that here the index contraction is equivalent to that of Eq. (\ref{aliceflies1}), enhanced with the conjugation of some components, which remain as an informative feature from the directionality of language, which would be lost otherwise. The circuit that calculates Eq. (\ref{alicefliesfinal}) is given in Fig. \ref{circ3}.

An alternative way of contracting the two-qubit spaces has been proposed in Ref. \cite{coecke2020foundations}, where the Bell effect $\bra{\beta_{00}} \equiv \left(\bra{00} + \bra{11}\right) \slash \sqrt{2} \in \mathcal{H}_N^1 \,  \otimes \,  \mathcal{H}_N^2$ is measured instead as

\begin{align}\label{bellcontraction}
 &\llbracket \text{Alice talks} \rrbracket = \bra{\beta_{00}} \ket{\text{Alice}} \ket{\text{talks}} \nonumber \\
 = & \frac{1}{\sqrt{2}} \sum_{p'q'r'} \left(\bra{00} + \bra{11} \right) A_{p'} t_{q'r'} \ket{n_{p'} n_{q'} s_{r'}} \nonumber \\
 = & \frac{1}{\sqrt{2}}\sum_{r'}  \left( A_{0} t_{0r'} + A_{1} t_{1r'} \right) \ket{s_{r'}}.
\end{align} Measuring the permutation operator as we do in Eq. (\ref{alicefliescalc}) is manifestly a more general way of contracting the representations of words than what is done in Eq. (\ref{bellcontraction}). On the one hand, it allows each interpretation space to have more than two basis states, that is, each quantum wire can represent something more general than one qubit. On the other hand, it accommodates correctly the existence of complex numbers in the quantum mechanical representations. Importantly, it has also one more important feature that we will make use of now: it allows us to integrate a quantum superposition of conflicting readings.

\subsection{Ambiguous readings on a quantum circuit}

\begin{figure}[!b]
\[ \Qcircuit @C=1em @R=2em @!R {
     & \lstick{1} & \qw & \qw & \qw &\qw\\
     & \lstick{2} & \qw & \multimeasure{3}{\parbox{0.7cm}{\centering $\xy ="j","j"-<.778em,.322em>;{"j"+<.778em,-.322em> \ellipse ur,_{}},"j"-<0em,.4em>;p+<.5em,.9em> **\dir{-},"j"+<1.4em,0.2em>*{},"j"-<1em,1em>*{} \endxy$\\$\widehat{P}_{25}$}} \inputgroupv{1}{2}{1.2em}{1.4em}{\ket{\text{rigorous}} \; \qquad \quad} \\
     & \lstick{\ket{\text{maths.}} \; 3} & \multimeasure{1}{\parbox{0.7cm}{\centering $\xy ="j","j"-<.778em,.322em>;{"j"+<.778em,-.322em> \ellipse ur,_{}},"j"-<0em,.4em>;p+<.5em,.9em> **\dir{-},"j"+<1.4em,0.2em>*{},"j"-<1em,1em>*{} \endxy$\\$\widehat{P}_{34}$}} &\\
     & \lstick{4} & \ghost{\parbox{0.7cm}{\centering $\xy ="j","j"-<.778em,.322em>;{"j"+<.778em,-.322em> \ellipse ur,_{}},"j"-<0em,.4em>;p+<.5em,.9em> **\dir{-},"j"+<1.4em,0.2em>*{},"j"-<1em,1em>*{} \endxy$\\$\widehat{P}_{34}$}} &  \\
     & \lstick{5} & \qw & \ghost{\parbox{0.7cm}{\centering $\xy ="j","j"-<.778em,.322em>;{"j"+<.778em,-.322em> \ellipse ur,_{}},"j"-<0em,.4em>;p+<.5em,.9em> **\dir{-},"j"+<1.4em,0.2em>*{},"j"-<1em,1em>*{} \endxy$\\$\widehat{P}_{25}$}} \\
     & \lstick{6} & \multimeasure{1}{\parbox{0.7cm}{\centering $\xy ="j","j"-<.778em,.322em>;{"j"+<.778em,-.322em> \ellipse ur,_{}},"j"-<0em,.4em>;p+<.5em,.9em> **\dir{-},"j"+<1.4em,0.2em>*{},"j"-<1em,1em>*{} \endxy$\\$\widehat{P}_{67}$}}   \inputgroupv{4}{6}{1.2em}{3em}{\ket{\text{and}} \; \quad } \\
     & \lstick{\ket{\text{physicists}} \;  7} & \ghost{\parbox{0.7cm}{\centering $\xy ="j","j"-<.778em,.322em>;{"j"+<.778em,-.322em> \ellipse ur,_{}},"j"-<0em,.4em>;p+<.5em,.9em> **\dir{-},"j"+<1.4em,0.2em>*{},"j"-<1em,1em>*{} \endxy$\\$\widehat{P}_{67}$}} & 
}\]
\caption{Quantum circuit for the first interpretation of the syntactically ambiguous phrase.}\label{circ4}
\end{figure}
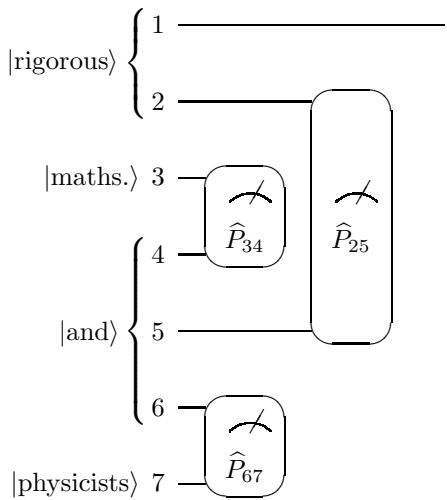

\begin{figure}[!b]
\[\Qcircuit @C=1em @R=2em @!R {
     & \lstick{1} & \qw & \multimeasure{3}{\parbox{0.7cm}{\centering $\xy ="j","j"-<.778em,.322em>;{"j"+<.778em,-.322em> \ellipse ur,_{}},"j"-<0em,.4em>;p+<.5em,.9em> **\dir{-},"j"+<1.4em,0.2em>*{},"j"-<1em,1em>*{} \endxy$\\$\widehat{P}_{14}$}} & \\
     & \lstick{2} & \multimeasure{1}{\parbox{0.7cm}{\centering $\xy ="j","j"-<.778em,.322em>;{"j"+<.778em,-.322em> \ellipse ur,_{}},"j"-<0em,.4em>;p+<.5em,.9em> **\dir{-},"j"+<1.4em,0.2em>*{},"j"-<1em,1em>*{} \endxy$\\$\widehat{P}_{23}$}}   \inputgroupv{1}{2}{1.2em}{1.4em}{\ket{\text{rigorous}} \; \qquad \quad} \\
     & \lstick{\ket{\text{maths.}} \; 3} & \ghost{\parbox{0.7cm}{\centering $\xy ="j","j"-<.778em,.322em>;{"j"+<.778em,-.322em> \ellipse ur,_{}},"j"-<0em,.4em>;p+<.5em,.9em> **\dir{-},"j"+<1.4em,0.2em>*{},"j"-<1em,1em>*{} \endxy$\\$\widehat{P}_{23}$}} &\\
     & \lstick{4} & \qw & \ghost{\parbox{0.7cm}{\centering $\xy ="j","j"-<.778em,.322em>;{"j"+<.778em,-.322em> \ellipse ur,_{}},"j"-<0em,.4em>;p+<.5em,.9em> **\dir{-},"j"+<1.4em,0.2em>*{},"j"-<1em,1em>*{} \endxy$\\$\widehat{P}_{14}$}}   \\
     & \lstick{5} & \qw & \qw & \qw & \qw\\
     & \lstick{6} & \multimeasure{1}{\parbox{0.7cm}{\centering $\xy ="j","j"-<.778em,.322em>;{"j"+<.778em,-.322em> \ellipse ur,_{}},"j"-<0em,.4em>;p+<.5em,.9em> **\dir{-},"j"+<1.4em,0.2em>*{},"j"-<1em,1em>*{} \endxy$\\$\widehat{P}_{67}$}}   \inputgroupv{4}{6}{1.2em}{3em}{\ket{\text{and}} \; \quad } \\
     & \lstick{\ket{\text{physicists}} \;  7} & \ghost{\parbox{0.7cm}{\centering $\xy ="j","j"-<.778em,.322em>;{"j"+<.778em,-.322em> \ellipse ur,_{}},"j"-<0em,.4em>;p+<.5em,.9em> **\dir{-},"j"+<1.4em,0.2em>*{},"j"-<1em,1em>*{} \endxy$\\$\widehat{P}_{67}$}} & 
}\] 
\caption{Quantum circuit for the second reading of the syntactically ambiguous phrase.}\label{circ5}
\end{figure}
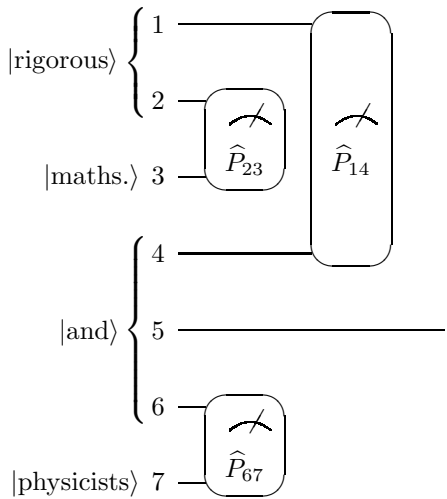

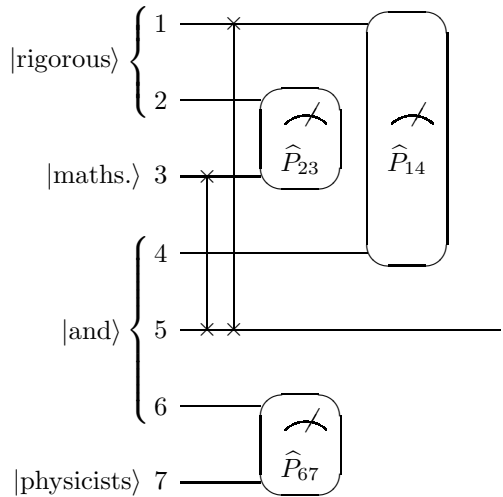
\begin{figure}[!tbh]
\[\Qcircuit @C=1em @R=2em @!R {
& \lstick{1} & \qw & \qswap \qwx[4] & \qw  & \multimeasure{3}{\parbox{0.7cm}{\centering $\xy ="j","j"-<.778em,.322em>;{"j"+<.778em,-.322em> \ellipse ur,_{}},"j"-<0em,.4em>;p+<.5em,.9em> **\dir{-},"j"+<1.4em,0.2em>*{},"j"-<1em,1em>*{} \endxy$\\$\widehat{P}_{14}$}} & \\
     & \lstick{2} & \qw & \qw  &  \multimeasure{1}{\parbox{0.7cm}{\centering $\xy ="j","j"-<.778em,.322em>;{"j"+<.778em,-.322em> \ellipse ur,_{}},"j"-<0em,.4em>;p+<.5em,.9em> **\dir{-},"j"+<1.4em,0.2em>*{},"j"-<1em,1em>*{} \endxy$\\$\widehat{P}_{23}$}}   \inputgroupv{1}{2}{1.2em}{1.4em}{\ket{\text{rigorous}} \; \qquad \quad} \\
     & \lstick{\ket{\text{maths.}} \; 3} & \qswap & \qw & \ghost{\parbox{0.7cm}{\centering $\xy ="j","j"-<.778em,.322em>;{"j"+<.778em,-.322em> \ellipse ur,_{}},"j"-<0em,.4em>;p+<.5em,.9em> **\dir{-},"j"+<1.4em,0.2em>*{},"j"-<1em,1em>*{} \endxy$\\$\widehat{P}_{23}$}} &\\
     & \lstick{4} & \qw \qwx & \qw  & \qw & \ghost{\parbox{0.7cm}{\centering $\xy ="j","j"-<.778em,.322em>;{"j"+<.778em,-.322em> \ellipse ur,_{}},"j"-<0em,.4em>;p+<.5em,.9em> **\dir{-},"j"+<1.4em,0.2em>*{},"j"-<1em,1em>*{} \endxy$\\$\widehat{P}_{14}$}}   \\
     & \lstick{5} & \qswap \qwx & \qswap  & \qw & \qw & \qw & \qw\\
     & \lstick{6} & \qw & \qw & \multimeasure{1}{\parbox{0.7cm}{\centering $\xy ="j","j"-<.778em,.322em>;{"j"+<.778em,-.322em> \ellipse ur,_{}},"j"-<0em,.4em>;p+<.5em,.9em> **\dir{-},"j"+<1.4em,0.2em>*{},"j"-<1em,1em>*{} \endxy$\\$\widehat{P}_{67}$}}   \inputgroupv{4}{6}{1.2em}{3em}{\ket{\text{and}} \; \quad } \\
     & \lstick{\ket{\text{physicists}} \;  7} & \qw & \qw & \ghost{\parbox{0.7cm}{\centering $\xy ="j","j"-<.778em,.322em>;{"j"+<.778em,-.322em> \ellipse ur,_{}},"j"-<0em,.4em>;p+<.5em,.9em> **\dir{-},"j"+<1.4em,0.2em>*{},"j"-<1em,1em>*{} \endxy$\\$\widehat{P}_{67}$}} & 
}\]
\caption{Quantum circuit that computes the first reading from the contractions of the second, and is therefore equivalent to Fig. \ref{circ4}.} \label{circ6}
\end{figure}

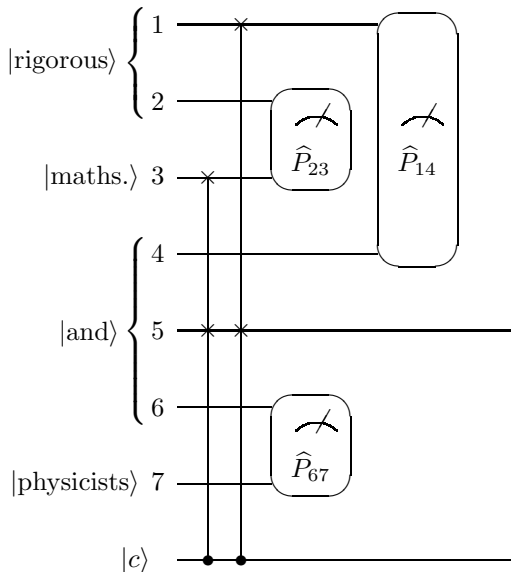
\begin{figure}[!tbh]
\[\Qcircuit @C=1em @R=2em @!R {
     & \lstick{1} & \qw & \qswap \qwx[6] & \qw  & \multimeasure{3}{\parbox{0.7cm}{\centering $\xy ="j","j"-<.778em,.322em>;{"j"+<.778em,-.322em> \ellipse ur,_{}},"j"-<0em,.4em>;p+<.5em,.9em> **\dir{-},"j"+<1.4em,0.2em>*{},"j"-<1em,1em>*{} \endxy$\\$\widehat{P}_{14}$}} & \\
     & \lstick{2} & \qw & \qw  &  \multimeasure{1}{\parbox{0.7cm}{\centering $\xy ="j","j"-<.778em,.322em>;{"j"+<.778em,-.322em> \ellipse ur,_{}},"j"-<0em,.4em>;p+<.5em,.9em> **\dir{-},"j"+<1.4em,0.2em>*{},"j"-<1em,1em>*{} \endxy$\\$\widehat{P}_{23}$}}   \inputgroupv{1}{2}{1.2em}{1.4em}{\ket{\text{rigorous}} \; \qquad \quad} \\
     & \lstick{\ket{\text{maths.}} \; 3} & \qswap \qwx[4] & \qw & \ghost{\parbox{0.7cm}{\centering $\xy ="j","j"-<.778em,.322em>;{"j"+<.778em,-.322em> \ellipse ur,_{}},"j"-<0em,.4em>;p+<.5em,.9em> **\dir{-},"j"+<1.4em,0.2em>*{},"j"-<1em,1em>*{} \endxy$\\$\widehat{P}_{23}$}} &\\
     & \lstick{4} & \qw  & \qw  & \qw & \ghost{\parbox{0.7cm}{\centering $\xy ="j","j"-<.778em,.322em>;{"j"+<.778em,-.322em> \ellipse ur,_{}},"j"-<0em,.4em>;p+<.5em,.9em> **\dir{-},"j"+<1.4em,0.2em>*{},"j"-<1em,1em>*{} \endxy$\\$\widehat{P}_{14}$}}   \\
     & \lstick{5} & \qswap  & \qswap  & \qw & \qw & \qw & \qw\\
     & \lstick{6} & \qw & \qw & \multimeasure{1}{\parbox{0.7cm}{\centering $\xy ="j","j"-<.778em,.322em>;{"j"+<.778em,-.322em> \ellipse ur,_{}},"j"-<0em,.4em>;p+<.5em,.9em> **\dir{-},"j"+<1.4em,0.2em>*{},"j"-<1em,1em>*{} \endxy$\\$\widehat{P}_{67}$}}   \inputgroupv{4}{6}{1.2em}{3em}{\ket{\text{and}} \; \quad } \\
     & \lstick{\ket{\text{physicists}} \;  7} & \qw  & \qw & \ghost{\parbox{0.7cm}{\centering $\xy ="j","j"-<.778em,.322em>;{"j"+<.778em,-.322em> \ellipse ur,_{}},"j"-<0em,.4em>;p+<.5em,.9em> **\dir{-},"j"+<1.4em,0.2em>*{},"j"-<1em,1em>*{} \endxy$\\$\widehat{P}_{67}$}} & \\
      & \lstick{\ket{c} \; \; } & \ctrl{-1} & \ctrl{-1} & \qw & \qw & \qw & \qw
}\]
\caption{Quantum circuit that computes simultaneously the two readings from the ambiguous input.}\label{circextra}
\end{figure}

The quantum states of the two readings in Eqs. (\ref{firstreadingfinal}) and (\ref{secondreadingfinal}), that result from contracting the individual word states, can be written as
\begin{align}\label{firstreadingq}
 &   \llbracket \text{rigorous mathematicians and physicists} \rrbracket_1 = \sum_{ijln} \, r_{ij} \, m_{l} \, a^*_{ljn} \, p^*_{n} \; \ket{{n}_i}
\end{align} and
\begin{align}\label{secondreadingq}
&   \llbracket \text{rigorous mathematicians and physicists} \rrbracket_2  = \sum_{jlmn} \, r^*_{lj} \, m_{j} \, a_{lmn} \, p^*_{n} \; \ket{{n}_m}.
\end{align} These can be represented by the two different circuits in Figs. \ref{circ4} and \ref{circ5} respectively, coming from the two different contraction schemes, as obtained in Ref. \cite{correia2020density}. Also in this reference, an analysis of how to express the two readings syntactic ambiguities simultaneously is developed, which we here implement. We can go from the first reading to the second by applying two wire swappings. First, we swap wires $3$ and $5$ on the second reading, which turns the measurement of $\widehat{P}_{23}$ into the measurement of $\widehat{P}_{25}$. Next, by swapping wires $5$ (which now contains the information from wire $3$) and $1$, we effectively turn the measurement of $\widehat{P}_{14}$ into the measurement of $\widehat{P}_{34}$. In this way, the circuit in Fig. \ref{circ6} is equivalent to that of the first reading. If we control the application of this set of swap gates on an extra qubit $\ket{c}=c_1\ket{1} + c_2\ket{0}$, we entangle the states of this qubit with the two possible readings. The first reading is stored in the quantum wire $5$ with probability $\abs{c_1}^2$, while the second reading is stored in that same quantum wire with probability $\abs{c_2}^2$. In total we have what is represented in the circuit of Fig. \ref{circextra}.

 The innovation that this implementation brings is that we are now able to deal with both interpretations simultaneously, that later contractions, with the representations of other words, have the potential to disambiguate.

\section{Application}\label{application}

In this section we apply Grover's quantum search algorithm to obtain the answer to a \textit{wh}-question with quantum speedup, using the quantum circuits for sentence representation developed in the previous section.

\subsection{Grover's quantum search algorithm}

Grover's quantum search algorithm aims at finding the correct answer to a query, by taking a state with an equal superposition of orthogonal states representing the answers as the input, and outputting a state in which the only basis states that have any probability of being measured correspond to correct answers. For $P=2^p$ possible solutions, the first step is to generate a linear superposition of $P$ unique states, with its index $x$ corresponding to one of the possible solutions. In the original proposal \cite{grover1996fast}, this input state is obtained by acting on $\ket{0}^{\otimes p}$ qubit states with the $H^{\otimes p}$ gate, where $H$ is the one-qubit Hadamard gate, which generates

\begin{equation}\label{multiplesol}
\ket{\Psi}=\frac{1}{\sqrt{P}} \sum_{x=0}^{P-1} \ket{x}.
\end{equation} Then, a sequence of gates, the \textit{Grover iteration}, is repeatedly applied to this input state, until a correct answer is the guaranteed outcome of the measurement of the initial qubits. For $Q$ correct solutions, only $O(\sqrt{P/Q})$ iterations are necessary, representing a quadratic speedup compared to a classical search algorithm, which requires checking all $P$ possible answers. Each Grover iteration $G$ has two main components: first, an \textit{oracle} operation $O$, and then an \textit{inversion about the mean} operation, formed by applying the unitary transformation that generates $\ket{\Psi}$ to $2\ket{0}\bra{0} - 1$, in this case

\begin{equation} 
H^{\otimes p} (2\ket{0}\bra{0} - 1)H^{\otimes p} \equiv 2\ket{\Psi}\bra{\Psi} -1,
\end{equation} which can easily be shown to be unitary. The heart of the algorithm is the oracle, as it is able to distinguish the answers that are correct from those that are not. It is a unitary operation that works by flipping the sign of the answer state if $\ket{x}$ is correct, that is,

\begin{flalign}
\left\{\begin{matrix}
& O(\ket{x})=-\ket{x} & \text{if} \ket{x} \text{is a correct answer,} \nonumber \\ 
& O(\ket{x})=\ket{x} & \text{otherwise}.
\end{matrix}\right. \nonumber
\end{flalign} To achieve this, more qubits might be necessary, and those constitute the ``oracle workspace". In the original setup, it is the oracle that depends on the query at hand, as well as the form of the inputs, while the inverse operation has a universal form. By representing our \textit{wh}-question query as quantum states and contractions therein, we will see that we can use Grover's algorithm with the problem-dependence of its parts reversed: instead it is the oracle that is universal, and the rotation is query-dependent, obtained from a unitary transformation on $\ket{0}$.

\subsection{Input-state preparation for question answering}

The question statement and possible answers hold the key for the search algorithm to identify the correct answers in our application. This will happen as a consequence of the contractions of the possible solutions with the question predicate. We will use our previous construction as the input of the first Grover iteration. To this end, suppose that we want to know the answer to the question \textit{Who talks?}, and that we have $P$ possible answers, of which $Q$ are absolutely correct, and $P-Q$ are, on the contrary, definitely wrong. For the oracle to identify the correct answers, they must be produced from the contraction with the verb, and they must be in a superposition equivalent to Eq. (\ref{multiplesol}).

The more complex mapping of $w$ to the semantics, when compared with the syntactic types $s$ and $n$, can be attributed to the particular semantics of questions and our application. 
In standard terms, the meaning of a question is taken as the map that sends answers, which belong to the interpretation space of nouns, to truth values, which are elements of the interpretation space of declarative sentences. We want to keep track of which word provides a correct answer in our quantum circuit, and a map like the latter, upon performing contractions, would only give us a count of how many correct and wrong answers there are. To see this, suppose that the word \textit{who} is represented in the space
$$\lceil w \slash (n\backslash s) \rceil =  \mathcal{H}_1^N \otimes \mathcal{H}_2^S \otimes \mathcal{H}_3^S \otimes \mathcal{H}_4^N,$$
and semantic representation as
\begin{align}
& \ket{\text{who}} =\sum_{ab,ij,kl}  \ket{n_i}_1  \ket{s_j}_2   \ket{s_k}_3 \ket{n_l}_4  \delta_{il} \delta_{jk} \nonumber \\
& = \sum_{i,l} \ket{n_i}_1 \ket{s_j}_2 \ket{s_j}_3 \ket{n_i}_4 .
\end{align}
and the intransitive verb \textit{talks} 
\begin{equation}
\ket{\text{talks}}= \sum_{mn} t_{mn} \; \ket{n_m}_5 \ket{s_n}_6.
\end{equation}
Following the contraction schemes presented in Sec.  \ref{demonstration}, the contraction between these two words states results in the state
\begin{equation}
\ket{\text{who}} \ket{\text{talks}}= \sum_{ij} t_{ij} \; \ket{n_i}_1 \ket{s_j}_2,
\end{equation} and representing the answers as
\begin{equation}
\ket{\text{answers}} = \sum_{p} W_{p}  \ket{n_p}_{7},
\end{equation} their final contraction results in
\begin{equation}\ket{\text{who}} \ket{\text{talks}} \ket{\text{answers}}= \sum_{ij} W^*_i t^*_{ij} \;\ket{s_j}_2.
\end{equation} This shows that a contraction just on the $S$ and $N$ spaces gives only a count of how many correct or incorrect answers there are, but not which ones are which.

As such, the map of the \textit{wh}-word needs to be furthermore tensored with elements of $\mathcal{H}^{\otimes p}$, of which each of the basis elements corresponds to the unique indexing of the possible answers. This provides an entanglement between the distributional representation of a noun, its corresponding truth value and an enumerable representation in the quantum circuit. The word \textit{who} thus belongs to the following semantic space, in the image of Eq. (\ref{whoref}),

$$ \lceil w \slash (n\backslash s) \rceil = \mathcal{H}_1^{\otimes p} \otimes \mathcal{H}_2^N \otimes \mathcal{H}_3^{\otimes p} \otimes \mathcal{H}_4^S \otimes \mathcal{H}_5^S \otimes \mathcal{H}_6^N,$$ with semantic representation given as 

\begin{align}
& \ket{\text{who}} =\sum_{ab,ij,kl} \ket{a}_1 \ket{n_i}_2 \ket{b}_3 \ket{s_j}_4   \ket{s_k}_5 \ket{n_l}_6 \delta_{ab} \delta_{il} \delta_{jk} \nonumber \\
& = \sum_{a,i,l} \ket{a}_1 \ket{n_i}_2  \ket{a}_3 \ket{s_j}_4 \ket{s_j}_5  \ket{n_i}_6 .
\end{align} The intransitive verb \textit{talks} has the same representation as before, now with the adapted labeling of wires

\begin{equation}
\ket{\text{talks}}= \sum_{mn} t_{mn} \; \ket{n_m}_7 \ket{s_n}_8.
\end{equation} For clarity, we flesh out the computation of the contractions that involve the extra index space. The semantic contraction of \textit{who} with \textit{talks}, following the interpretation of the syntactic contraction, results in 

\begin{align}\label{whoflies}
&\bra{\text{who}} \bra{\text{talks}} \; \widehat{P}_{67} \otimes \widehat{P}_{58} \otimes \widehat{O}_{1234} \; \ket{\text{who}} \ket{\text{talks}} \nonumber\\
&     =  \sum_{a'i'j'm'n'} \bra{a'}_1 \bra{n_{i'}}_2  \bra{a'}_3 \bra{s_{j'}}_4   \bra{s_{j'}}_5 \bra{n_{i'}}_6 t^*_{m'n'} \bra{n_{m'}}_7 \bra{s_{n'}}_8 \nonumber \\
& \cdot \widehat{O}_{1234} \; \sum_{aijmn}  \ket{a}_1 \ket{n_{i}}_2  \ket{a}_3 \ket{s_{j}}_4 \ket{s_{n}}_5  \ket{n_{m}}_6   t_{mn} \ket{n_{i}}_7 \ket{s_{j}}_8 \nonumber \\
    &  =  \sum_{a'i'j'm'n'} \bra{a'}_1 \bra{n_{i'}}_2  \bra{a'}_3 \bra{s_{j'}}_4    t^*_{m'n'}  \widehat{O}_{1234} \;   \ket{a}_1 \ket{n_{m'}}_2  \ket{a}_3 \ket{s_{n'}}_4    t_{i'j'}  \nonumber \\
     & = \sum_{a'i'j'} t_{i'j'} \bra{a'}_1 \bra{n_{i'}}_2  \bra{a'}_3 \bra{s_{j'}}_4    \widehat{O}_{1234} \; \sum_{ail} t^*_{ij}   \ket{a}_1 \ket{n_{i}}_2  \ket{a}_3 \ket{s_{j}}_4.    
\end{align} From this we read off the question representation, rewriting the indices:

\begin{equation}
    \llbracket \text{who talks} \rrbracket = \sum_{aij} t^*_{ij}   \ket{a}_1 \ket{n_{i}}_2  \ket{a}_3 \ket{s_{j}}_4.
\end{equation}

\begin{figure}[!tbh]
\[\Qcircuit @C=1em @R=2em @!R {
&\lstick{1} & \qw & \qw & \qw & \multimeasure{9}{\parbox{0.7cm}{\centering $\xy ="j","j"-<.778em,.322em>;{"j"+<.778em,-.322em> \ellipse ur,_{}},"j"-<0em,.4em>;p+<.5em,.9em> **\dir{-},"j"+<1.4em,0.2em>*{},"j"-<1em,1em>*{} \endxy$\\$\widehat{P}_{1,10}$}} \\
&\lstick{2} & \qw & \qw & \multimeasure{7}{\parbox{0.7cm}{\centering $\xy ="j","j"-<.778em,.322em>;{"j"+<.778em,-.322em> \ellipse ur,_{}},"j"-<0em,.4em>;p+<.5em,.9em> **\dir{-},"j"+<1.4em,0.2em>*{},"j"-<1em,1em>*{} \endxy$\\$\widehat{P}_{29}$}} \\
&\lstick{3} & \qw & \qw & \qw & \qw & \qw & \qwa \\
&\lstick{4} & \qw & \qw & \qw & \qw & \qw & \qwa \\
&\lstick{5} & \qw & \multimeasure{3}{\parbox{0.7cm}{\centering $\xy ="j","j"-<.778em,.322em>;{"j"+<.778em,-.322em> \ellipse ur,_{}},"j"-<0em,.4em>;p+<.5em,.9em> **\dir{-},"j"+<1.4em,0.2em>*{},"j"-<1em,1em>*{} \endxy$\\$\widehat{P}_{58}$}} \\
&\lstick{6} & \multimeasure{1}{\parbox{0.7cm}{\centering $\xy ="j","j"-<.778em,.322em>;{"j"+<.778em,-.322em> \ellipse ur,_{}},"j"-<0em,.4em>;p+<.5em,.9em> **\dir{-},"j"+<1.4em,0.2em>*{},"j"-<1em,1em>*{} \endxy$\\$\widehat{P}_{67}$}}\\
&\lstick{7} & \ghost{\parbox{0.7cm}{\centering $\xy ="j","j"-<.778em,.322em>;{"j"+<.778em,-.322em> \ellipse ur,_{}},"j"-<0em,.4em>;p+<.5em,.9em> **\dir{-},"j"+<1.4em,0.2em>*{},"j"-<1em,1em>*{} \endxy$\\$\widehat{P}_{67}$}} \\
&\lstick{8} & \qw & \ghost{\parbox{0.7cm}{\centering $\xy ="j","j"-<.778em,.322em>;{"j"+<.778em,-.322em> \ellipse ur,_{}},"j"-<0em,.4em>;p+<.5em,.9em> **\dir{-},"j"+<1.4em,0.2em>*{},"j"-<1em,1em>*{} \endxy$\\$\widehat{P}_{58}$}} \\
&\lstick{9} & \qw & \qw & \ghost{\parbox{0.7cm}{\centering $\xy ="j","j"-<.778em,.322em>;{"j"+<.778em,-.322em> \ellipse ur,_{}},"j"-<0em,.4em>;p+<.5em,.9em> **\dir{-},"j"+<1.4em,0.2em>*{},"j"-<1em,1em>*{} \endxy$\\$\widehat{P}_{29}$}} \\
&\lstick{\quad 10} & \qw & \qw & \qw  &  \ghost{\parbox{0.7cm}{\centering $\xy ="j","j"-<.778em,.322em>;{"j"+<.778em,-.322em> \ellipse ur,_{}},"j"-<0em,.4em>;p+<.5em,.9em> **\dir{-},"j"+<1.4em,0.2em>*{},"j"-<1em,1em>*{} \endxy$\\$\widehat{P}_{1,10}$}}
\inputgroupv{1}{6}{1.5em}{7.2em}{\ket{\text{who}} \qquad}
\inputgroupv{7}{8}{1.5em}{1.3em}{\ket{\text{talks}} \qquad}
\inputgroupv{9}{10}{1.5em}{1.3em}{\ket{\text{answers}} \qquad \quad \;}
}\]
\caption{Quantum circuit that generates the input of the first Grover iteration for question-answering.} \label{circ7}
\end{figure}
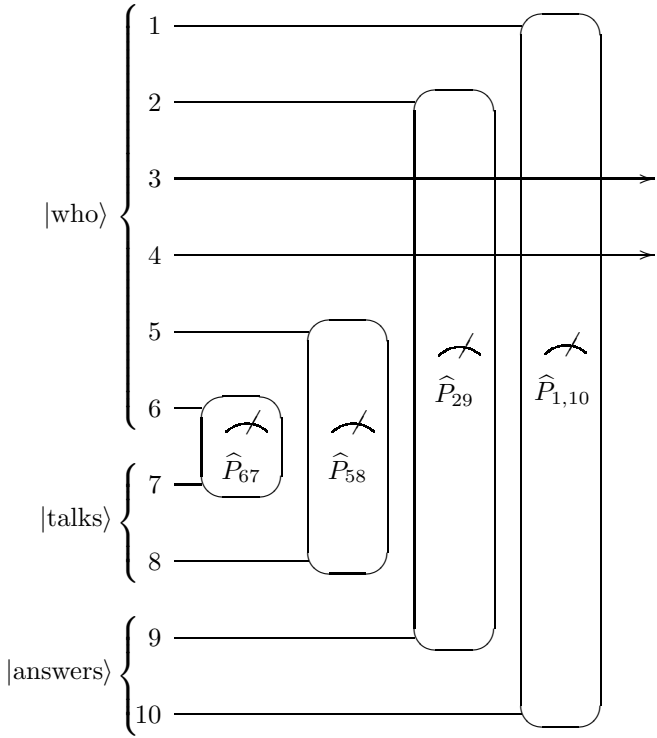

To be the input of the quantum search algorithm, the full input needs to correspond to an equal superposition of all possible answers. This can be achieved by entangling the distributional representation of the answers with the corresponding index

\begin{equation}
\ket{\text{answers}} = \sum_{bp} W_{p}^{b}  \ket{n_p}_{9} \ket{b}_{10}, 
\end{equation} followed by a contraction with the question representation, which happens strictly at the semantic level. Hence

\begin{align}
&\bra{\text{who talks}} \bra{\text{answers}} \widehat{P}_{29} \otimes \widehat{P}_{1,10} \otimes \widehat{O}_{34}  \ket{\text{who talks}} \ket{\text{answers}}=  \nonumber \\
& \sum_{a'i'l'} t_{i'j'} \bra{a'}_1 \bra{n_{i'}}_2  \bra{a'}_3 \bra{s_{j'}}_4 \sum_{b'p'} W_{p'}^{*b'} \bra{n_{p'}}_{9} \bra{b'}_{10}  \nonumber \\
& \cdot \widehat{O}_{34} \sum_{ail} t^*_{ij}   \ket{b}_1 \ket{n_{p}}_2  \ket{a}_3 \ket{s_{j}}_4 \sum_{bp} W_p^b \ket{n_{i}}_{9} \ket{a}_{10}  \nonumber \\
& = \sum_{a'i'j'} W^{a'}_{i'} t_{i'j'}      \bra{a'}_3 \bra{s_{j'}}_4 \widehat{O}_{34} \sum_{amn} W^{* a}_{ i} t^*_{ij}    \ket{a}_3 \ket{s_{j}}_4,
\end{align} such that the effective input state to the Grover's algorithm is

\begin{equation}\label{psi_initial}
\ket{\Psi_{initial}}=\sum_{aij} W^{*a}_{i} t^*_{ij}    \ket{a}_3 \ket{s_{j}}_4,
\end{equation} which represents an entanglement between the indices of possible answers and truth values. This process is represented by the circuit in Fig. \ref{circ7}.


\subsection{Oracle and inversion}

\begin{figure}[!tbh]
\[\Qcircuit @C=1em @R=2em @!R {
     & \lstick{3} & \qw & \qw & \qw & \qw & \qw & \qwa  \\
     & \lstick{4} & \qw & \qw  & \qw & \ctrl{1}  & \qw &  \qwa  \inputgroupv{1}{2}{1.2em}{1.5em}{\ket{\Psi_{initial}} \qquad \quad } \\
     & \lstick{\ket{0} \;  11} & \gate{H} & \gate{X} & \qw & \targ & \qw  & \qwa
}\]
\caption{Oracle for question-answering.}\label{circ8}
\end{figure}
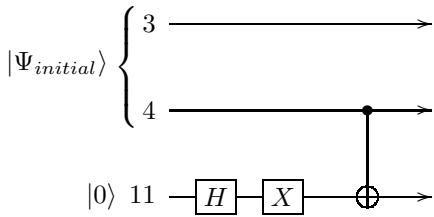

Grover's algorithm requires a normalized state as initial input. Since the amplitudes in the state in Eq. (\ref{psi_initial}) have information about whether a certain answer is correct, they uniquely associate each word indexed by $a$ with one of the basis states $\ket{s_{j}}$, in such a way that $\sum_i W^{*a}_{i} t^*_{ij}$  is null if the combination between word index $a$ and truth value $j$ is not correct, and otherwise equal to one. Since every word should be either true or false, that leaves us with precisely $P$ independent and equally summed states. Therefore, if the indices $aj$ are abbreviated by one index $x$, the normalized state is given as

\begin{equation}
\ket{\Psi_{initial}}  = \frac{1}{\sqrt{P}} \sum_{x=0}^{P-1} \ket{x},
\end{equation} with

\begin{equation}
    \ket{x}=\sum_{i} W^{*a}_{i} t^*_{ij}    \ket{a}_3 \ket{s_{j}}_4.
\end{equation} The oracle applied to this input state takes the form of the circuit in Fig. \ref{circ8}.

The states $\ket{a}$ on the first $p$ qubits, being in one-to-one correspondence with each of the possible solutions, are the complete set of states that build up the equal superposition $\ket{\Psi}=H^{\otimes p} \ket{0}^{\otimes p}$, as in Eq. (\ref{multiplesol}). In terms of the states that correspond to words that are correct or incorrect, we can rewrite $\ket{\Psi}$ as 

\begin{equation}
    \ket{\Psi}=\cos(\frac{\theta}{2})\ket{\alpha}_3 + \sin(\frac{\theta}{2})\ket{\beta}_3,
\end{equation} with $\ket{\alpha}$ the normalized sum of all states that correspond to words that are not solutions, and $\ket{\beta}$ to the normalized sum of those that correspond to words that are solutions. Using this notation, the $\ket{\Psi_{initial}}$ state that we obtain using the contractions can be expressed as

\begin{align}\label{anglestate}
&\ket{\Psi_{initial}} = \cos(\frac{\theta}{2})\ket{\alpha}_3\ket{0}_4 + \sin(\frac{\theta}{2})\ket{\beta}_3\ket{1}_4.
\end{align} Though there is entanglement between the first $p$ qubits and the last one, this is a pure state in the $p+1$ qubit space, as it results from the measurement of the permutation operators, as shown in Sec. \ref{demonstration}. As such, there is a unitary transformation that generates it from $\ket{0}^{\otimes p+1}$ as

\begin{equation}
    \ket{\Psi_{initial}}=U\ket{0}^{p+1}.
\end{equation} Using this, we can construct the rotation part of the Grover algorithm as 

\begin{equation}\label{rotation}
U(2\ket{0}\bra{0} - 1) U^{\dagger}= 2\ket{\Psi_{initial}}\bra{\Psi_{initial}} - 1.
\end{equation} It follows that the Grover iteration applied on the input state in Eq. (\ref{anglestate}) using the oracle and the rotation in Eq. (\ref{rotation}) gives the desired outcome of

\begin{align}
&(2\ket{\Psi_{initial}}\bra{\Psi_{initial}} - 1)O(\ket{\Psi_{initial}})= \nonumber \\ &\cos(\frac{3\theta}{2})\ket{\alpha}_3\ket{0}_4 + \sin(\frac{3\theta}{2})\ket{\beta}_3\ket{1}_4.
\end{align} This is precisely what we expect to obtain. For the second iteration, the oracle acts as desired, and so does the rotation. After a number of iterations only the states associated with $\ket{1}_4$ have positive amplitude, which means that we are certain to measure the index of a word that corresponds to a correct answer when we make a measurement on the first $p$ qubits. Thus we have obtained a correct answer with quadratic speedup due to the quantum search algorithm.

\section{Conclusion and Outlook}\label{conclusion}

In this paper we introduced two main developments in the application of quantum computation to natural language processing. The first one is a tensor-contraction scheme on quantum circuits. Taking quantum states as the representations of all input words, contractions are then identified with the expectation value of an appropriate permutation operator. Doing this, are we not only able to reproduce previous analytical results, but we also allow for complex values and create quantum circuits that are equipped to deal with the syntactic ambiguities in Ref. \cite{correia2020density}. With this setup, each reading of an ambiguous phrase corresponds to a particular circuit, and different readings are interchangeable upon the application of a number of swap gates. Controlling on these swap gates, we can obtain a true quantum superposition of the multiple readings. This covers the problem of how to deal with multiple readings in real time, without the need to assume any contextualization. While this addresses the question of syntactic ambiguities by making use of the quantum superposition principle, ambiguities at the word level can be immediately accommodated for by using density matrices \cite{piedeleu2015open,DBLP:journals/amai/SadrzadehKB18,bankova2019graded,meyer2020modelling}, instead of the pure states we use here for simplicity. A generalization to other sentence-level ambiguities constitutes further work, in the expectation that the use of different controls allows for different readings simultaneously in the output state. Note that, in terms of a concrete implementation, the permutation between two qubits used to generate an equal superposition of readings from an ambiguous input takes the form of a Fredkin gate, or a CSWAP gate, which might add considerable circuit complexity, but this is expected to be compensated by the fact that the number of two-qubit operations only scales linearly with an increasing number of readings, since for these types of ambiguities all permutations can be generated via sets of SWAP operations. We leave a more robust exploration of these technical constraints to future work.

The second development builds on this quantum framework, and consists of a quantum search algorithm that is able to find the answer to a \textit{wh}-question with quantum speedup. As input, the algorithm takes a multi-partite state in quantum superposition, representing a \textit{wh}-question and its possible answers, and performs a series of tensor contractions as established previously. A series of gates then acts repeatedly on the post-contraction state, guaranteeing that a correct answer to the question is obtained upon a single final measurement. Our algorithm takes advantage of intrinsic quantum features to identify and deliver a correct answer with quantum speedup of quadratic order, when compared to the classical alternative of checking every possible answer. We are thus able to provide a correct answer using information given directly by the tensor contractions of representations of words as proposed by DisCoCat, and without needing to hand-feed any labels nor to learn the answers to other questions. Our approach thus shows how quantum circuit implementations can break with the widely accepted ``learning paradigm" of current NLP approaches to question-answering and other tasks used in Ref. \cite{meichanetzidis2020grammar}, providing a scalable approach to open-ended questions. 
Our approach differs from that of Ref. \cite{coecke2019mathematics} also in the sense that we keep all words as input states, instead of representing words from complex types as gates that modify circuit inputs, remaining closer to the compositional spirit of the syntax and therefore being more easily extensible to larger language fragments. 
Further work includes finding an effective implementation of the measurement of the permutation operator for an arbitrary number of qubits, possibly making use of the Hadamard test \cite{aharonov2009polynomial}, and understanding how to find a universal form of the inversion operator that does not depend on $\ket{\alpha}$ and $\ket{\beta}$ separately. An extension of the present formalism can furthermore account for a better understanding of the temporal evolution of the meanings of sentences.

\section*{Acknowledgements}

We would like to thank the anonymous referees for the helpful comments. This work is supported by the Utrecht University's  Complex Systems Fund, with special thanks to Peter Koeze, and the D-ITP consortium, a program of the Netherlands Organization for Scientific Research (NWO) that is funded by the Dutch Ministry of Education, Culture and Science (OCW).

\section*{Data Availability Statement}

All data generated or analyzed during this study are included in this published article.

\bibliography{sn-bibliography}

\begin{appendices}

\section{Measuring the permutation operator on two qubits}\label{permutationappendix}

In this appendix we show that for the measurement of the permutation operator applied to two qubits it suffices to measure the input states in the Bell basis. The two input qubits have the four possible joint states in the standard basis, given by

\begin{align}
&\ket{00}= \begin{pmatrix}
1\\ 
0\\ 
0\\ 
0
\end{pmatrix}, \ket{10}= \begin{pmatrix}
0\\ 
1\\ 
0\\ 
0
\end{pmatrix}, \ket{01}= \begin{pmatrix}
0\\ 
0\\ 
1\\ 
0
\end{pmatrix}, \ket{11}= \begin{pmatrix}
0\\ 
0\\ 
0\\ 
1
\end{pmatrix}.
\end{align} The permutation operator applied to two qubits is equivalent to the SWAP gate $S$. In this basis, this operator has the matrix representation

\begin{equation}
S = \begin{pmatrix}
1 & 0 & 0 & 0\\ 
0 & 0 & 1 & 0\\ 
0 & 1 & 0 & 0\\ 
0 & 0 & 0 & 1
\end{pmatrix}.
\end{equation} The eigenstates of this operator are the well-known singlet and triplet states that represent the joint spin of two spin-$1/2$ particles. With eigenvalue $-1$, we have the singlet state

\begin{equation}
\ket{0,0}=\frac{1}{\sqrt{2}}    \begin{pmatrix}
0\\ 
-1\\ 
1\\ 
0
\end{pmatrix},
\end{equation} and with eigenvalue $1$ we have the three triplet states

\begin{align}
&\ket{1,-1}=   \begin{pmatrix}
1\\ 
0\\ 
0\\ 
0
\end{pmatrix}, 
\ket{1,0}=   \frac{1}{\sqrt{2}}    \begin{pmatrix}
0\\ 
1\\ 
1\\ 
0
\end{pmatrix},  \text{ and }   \ket{1,1}=  \begin{pmatrix}
0\\ 
0\\ 
0\\ 
1
\end{pmatrix},
\end{align} expressed in the standard basis as

\begin{align} 
& \ket{0,0}=\frac{1}{\sqrt{2}}(\ket{01} - \ket{10}) \label{singlet}, \\
& \ket{1,-1}=\ket{00}, \\
& \ket{1,0}= \frac{1}{\sqrt{2}}(\ket{01} + \ket{10}), \\
& \ket{1,1}=\ket{11}. \label{triplet}
\end{align} In its turn, the Bell basis can be expressed in terms of the standard basis in the following way

\begin{align}
& \ket{\beta_{00}}=\frac{1}{\sqrt{2}}(\ket{00} + \ket{11}), \\
&\ket{\beta_{01}}=\frac{1}{\sqrt{2}}(\ket{01} + \ket{10}), \\
& \ket{\beta_{10}}= \frac{1}{\sqrt{2}}(\ket{00} - \ket{11}), \\
& \ket{\beta_{11}}= \frac{1}{\sqrt{2}}(\ket{01} - \ket{10}).
\end{align} The Bell states can thus be rewritten using the total-spin eigenstates of $S$, given in (\ref{singlet}) to (\ref{triplet}), as:

\begin{align}
& \ket{\beta_{00}}=\frac{1}{\sqrt{2}}(\ket{1,-1} + \ket{1,1}) \\
&\ket{\beta_{01}}=\ket{1,0} \\
& \ket{\beta_{10}}= \frac{1}{\sqrt{2}}(\ket{1,-1} - \ket{1,1}) \\
& \ket{\beta_{11}}= \ket{0,0}.
\end{align} Because any linear combination of degenerate eigenstates is also an eigenstate of that operator with the same eigenvalue [proof: $A\vec{v}=\lambda \vec{v}, A\vec{w}=\lambda \vec{w} \Rightarrow A(a\vec{v}+b\vec{w})=\lambda(a\vec{v}+b\vec{w})$], we see that $\ket{\beta_{00}}$, $\ket{\beta_{01}}$ and $\ket{\beta_{10}}$ are eigenstates of $S$ with eigenvalue $1$, and $\ket{\beta_{11}}$ is an eigenstate with eigenvalue $-1$. Therefore, we can conclude that the Bell basis also diagonalizes the permutation operator, and as such repeated measurements of the qubits in this basis allows us to directly compute the expectation value of the operator in the input states. So, for a two-qubit input state

\begin{equation}
\ket{\Phi} = \sum_{ij} a_i b_j \ket{ij},
\end{equation} with $i,j \in \{0,1\}$, the expectation value of the $S$ operator is given by

\begin{align}
&\expval{S}_\Phi= \abs{ \bra{\beta_{00}} \ket{\Phi}} ^2 + \abs{ \bra{\beta_{01}} \ket{\Phi}} ^2 + \abs{ \bra{\beta_{10}} \ket{\Phi}} ^2 - \abs{ \bra{\beta_{11}} \ket{\Phi}}^2. 
\end{align} If we are in the possession of a measuring device that can only measure in the standard basis, we must transform our input states with the inverse transformation that generates the Bell states. This serves to guarantee that an outcome $\ket{ij}$ is in fact as likely as the measurement of $\ket{\beta_{ij}}$ if the input states were measured directly in the Bell basis.

\end{appendices}

\end{document}